\documentclass[journal=jctcce,manuscript=article]{achemso} % {traditional, twocolumn}

\usepackage[T1]{fontenc} % Use modern font encodings
\usepackage[version=3]{mhchem} % Formula subscripts using \ce{}
\usepackage[dvipsnames]{xcolor}
\usepackage{graphicx,subcaption}
\usepackage{amssymb}
\usepackage{overpic}
\usepackage{soul}
\usepackage{bm}

\newif\ifhighlightchanges
\highlightchangesfalse

\definecolor{darkblue}{HTML}{003D6D}

\newcommand{\bR}{{\bm R}}
\newcommand{\bP}{{\bm P}}

\newcommand{\bx}{{\bm x}}
\newcommand{\bv}{{\bm v}}
\newcommand{\bp}{{\bm p}}

\newcommand{\bd}{{\bm d}}

\newcommand{\bK}{{\bm K}}
\newcommand{\bX}{{\bm X}}
\newcommand{\bGamma}{{\bm \Gamma}}

\newcommand{\braket}[3]{\left< #1\left|#2\right|#3 \right>}

\newcommand{\bra}[1]{\left< #1\right|}
\newcommand{\ket}[1]{\left|#1 \right>}
\newcommand{\pp}[2]{\frac{\partial {#1}}{\partial {#2}}}

\author{Tian Qiu}
\affiliation{Department of Chemistry, University of Pennsylvania, Philadelphia, Pennsylvania 19104, USA.}
\author{Mansi Bhati}
\affiliation{Department of Chemistry, University of Pennsylvania, Philadelphia, Pennsylvania 19104, USA.}
\author{Zhen Tao}
\affiliation{Department of Chemistry, University of Pennsylvania, Philadelphia, Pennsylvania 19104, USA.}
\author{Xuezhi Bian}
\affiliation{Department of Chemistry, University of Pennsylvania, Philadelphia, Pennsylvania 19104, USA.}
\author{Jonathan Rawlinson}
\affiliation{Department of Mathematics, University of Manchester, Manchester M13 9PL, UK}
\author{Robert G. Littlejohn}
\affiliation{Department of Physics, University of California, Berkeley, California 94720, USA}
\author{Joseph E. Subotnik}
\email{subotnik@sas.upenn.edu}
\affiliation{Department of Chemistry, University of Pennsylvania, Philadelphia, Pennsylvania 19104, USA.}

\title[]{A Simple One-Electron Expression for Electron Rotational Factors}

\begin{document}

\begin{abstract}
Within the context of FSSH dynamics, one often wishes to remove the angular component of the derivative coupling between states $\ket{J}$ and $\ket{K}$. In a  set of previous papers, Truhlar {\em et al.} posited one approach for such a  removal based on direct projection, while we isolated a second approach  by constructing and differentiating rotationally invariant basis. Unfortunately, neither approach was able to demonstrate a {\em one-electron operator  $\hat{O}$} whose matrix element $\braket{J}{\hat{O}}{K}$ was the angular component of the derivative coupling. Here, we show that a one-electron operator can in fact be constructed efficiently in a semi-local fashion. The present results yield physical insight  into designing new surface hopping algorithms and be of immediate use for FSSH calculations.
\end{abstract}

\section{Introduction: Surface Hopping and Linear/Angular Momentum Conservation}
Surface hopping is today the most popular mixed quantum-classical algorithm for propagating nonadiabatic dynamics\cite{tretiak:2014:acr,furche:2013:fssh_review}, offering a reasonable balance between speed vs accuracy, while also  roughly recovering the correct equilibrium density distribution.\cite{tully:2005:detailedbalance,tully:2008:detailedbalance} The essence of the algorithm is to follow dynamics along adiabats, with occasional jumps between adiabats so as to account for electronic relaxation. Importantly, by propagating along adiabats, the algorithm automatically conserves the total energy.
That being said, the standard FSSH algorithm does not conserve linear momentum conservation\cite{fatehi:2011:dercouple}, a failure that has been addressed before in the literature; angular momentum is also not conserved\cite{truhlar:2020:angular,Athavale2023, Bian:2023}, though this problem is much less well appreciated and discussed (except in the context of exact factorization approaches\cite{gross:2010:exact_factorization,gross:2016:exact_factorization,gross:2022:epl_transfer} and Coriolis force\cite{Geilhufe:2022}).

Momentum conservation fails within FSSH because, when a trajectory hops between electronic states, the fundamental ansatz  of surface hopping is that the momentum rescaling (between states $\ket{J}$ and $\ket{K}$) should occur along the derivative coupling direction $\bd_{JK}$ between these two states,
\begin{align}
\label{ddef}
d_{JK}^{A\alpha} = \braket{J}{\frac{\partial}{\partial X_{A\alpha}}}{K},
\end{align}
Here and below, we use $A,B,C$ to index nuclei,   $I,J,K$ to index adiabatic electronic states, $\alpha,\beta,\gamma$ to index an $x,y,z$ Cartesian direction. (Although not present in Eq. \ref{ddef}, note also three dimensional vectors are written in {\bf bold} font; $\mu,\nu, \lambda, \sigma$ index atomic orbitals $\ket{\chi_\mu},\ket{\chi_\nu},\ket{\chi_\lambda},\ket{\chi_\sigma}$, respectively) Now, the nature of the derivative couplings as a function of translation and rotation has been studied in the past.\cite{fatehi:2011:dercouple}.  In short, when dealing with the standard electronic Hamiltonian (i.e. without spin-orbital coupling), the usual phase conventions\cite{Littlejohn:2023} are that the nuclei and electrons are translated together (so that the total wavefunction is real-valued). Mathematically, this means we choose the phase of state $\ket{K}$ to follow:
\begin{align}
    \left( \hat{\bm{P}}_e + \hat{\bm{P}}_N \right)\ket{K} = 0, \\
    \left( \hat{\bm{L}}_e + \hat{\bm{L}}_N \right)\ket{K} = 0,
\end{align}
where $\hat{\bm{P}}_e$ and $\hat{\bm{P}}_N$ are electronic and nuclear linear momentum operators, respectively; $\hat{\bm{L}}_e$ and $\hat{\bm{L}}_N$ are electronic and nuclear angular momentum operators, respectively. If one operates by $\bra{J}$, one then automatically finds that:
\begin{align}
    \braket{J}{\hat{\bm{P}}_e}{K} + \braket{J}{\hat{\bm{P}}_N}{K} &=0,\\
    \braket{J}{\hat{\bm{L}}_e}{K} + \braket{J}{\hat{\bm{L}}_N}{K} &=0.
\end{align}
These expressions can also be written as:
\begin{align}
    \braket{J}{\hat{\bm{P}}_e}{K} + \frac{\hbar}{i}\sum_A \braket{J} {\frac{\partial }{\partial \bX_{A}}}{K} &= \braket{J}{\hat{\bm{P}}_e}{K} + \frac{\hbar}{i}\sum_A \bd_{JK}^{A} = 0,\label{eq:PJK}\\
    \braket{J}{\hat{\bm{L}}_e}{K} + \frac{\hbar}{i} \sum_A \bX_A \times \braket{J}{\frac{\partial }{\partial \bX_{A}}}{K} &=    \braket{J}{\hat{\bm{L}}_e}{K} + \frac{\hbar}{i} \sum_A \bX_A \times \bd_{JK}^A =0,
\end{align}
where ``$\times$'' represents the cross product. Thus, at the end of the day, rescaling the classical nuclear momentum by ${\bm d}$,
\begin{align}
\bP^A_{\rm final} = \bP^A_{\rm initial} + \alpha \bd^A
\end{align}
must lead to a violation of linear conservation insofar as 
\begin{align}
    \sum_A \bP^A_{\rm final} = \sum_A \bP^A_{\rm initial} +i \hbar \alpha \braket{J}{\hat{\bm{P}}_e}{K} \ne \sum_A \bP^A_{\rm initial}
\end{align}
At bottom,  nuclear displacement  displaces the electrons (which yield a small change in total momentum). Similar statements hold for angular momentum.

For linear and angular momentum conservation,  the most natural approach is to  modify the rescaling direction by:\begin{align}
\bP^A_{\rm final} =  \bP^A_{\rm initial} + \alpha  (\bd^A -\bm{\Gamma}^A),
\end{align}
where $\bGamma$ satisfies:
\begin{align}
    \sum_A \bGamma_{JK}^{A} &=\frac{\bp_{JK}}{i\hbar}\label{eq:PP},\\
    \sum_{A}\bm{X}_{A}\times\bm{\Gamma}^{A}_{JK} &=\frac{\bm{l}_{JK}}{i\hbar}\label{eq:LL},
\end{align}
Here, $\bp_{JK}$ and $\bm{l}_{JK}$ are the electronic linear and angular momentum matix elements between states $\ket{J}$ and $\ket{K}$, respectively. In practice, one often decomposes
 $$\bm{\Gamma} = \bm{\Gamma}' + \bm{\Gamma}'',$$ 
 where $\bm{\Gamma}'$ is denoted an electron translation factor (ETF),  and $\bm{\Gamma}''$ is denoted an electron rotation factor (ERF).

 Now, as written above, the $\bGamma'$ and $\bGamma''$ tensors are matrices in a vector space composed of many-body
 electronic wavefunctions, $\ket{J},\ket{K}$, i.e. $\bGamma = \bGamma_{JK}$. In practice, working with such matrices is quite difficult and it would be much better if one could fashion these matrices as one-electron operators (in an atomic orbital basis) instead. In other words, rather than constructing $\bGamma'_{JK}$ and $\bGamma''_{JK}$ above, it would be extremely convenient if we could define operators 
 $\bGamma'_{\mu \nu}$ and $\bGamma''_{\mu \nu}$
and thereafter evaluate the matrices:
\begin{align}
    \bGamma'_{JK} =  \sum_{\mu \nu} \bGamma'_{\mu \nu} D_{\mu \nu}^{JK}, \label{eq:etf_jk}\\
    \bGamma''_{JK} =  \sum_{\mu \nu} \bGamma''_{\mu \nu} D_{\mu \nu}^{JK}.\label{eq:erf_jk}
\end{align}
Here, $D_{\mu \nu}^{JK}$ is the one-electron transition density matrix between states $\ket{J}$ and $\ket{K}$. Since $\bm{p}_{JK} = \sum_{\mu\nu}\bm{p}_{\mu\nu}D_{\mu \nu}^{JK}$ and $\bm{l}_{JK} = \sum_{\mu\nu}\bm{l}_{\mu\nu}D_{\mu \nu}^{JK}$, the simplest means to satisfy Eqs. \ref{eq:PP}-\ref{eq:LL} (given the definitions in  Eqs. \ref{eq:etf_jk}-\ref{eq:erf_jk}) is to require:

\begin{align}
    &\boxed{\sum_A\bm{\Gamma}_{\mu\nu}^A = \sum_A \left(\bm{\Gamma}'^A_{\mu\nu}+\bm{\Gamma}''^A_{\mu\nu}\right) = \frac{\bm{p}_{\mu\nu}}{i\hbar}}\label{eq:etf_munu}\\
    &\boxed{\sum_A\bm{X}_A\times\bm{\Gamma}_{\mu\nu}^A = \sum_A \bm{X}_A\times\left(\bm{\Gamma}'^A_{\mu\nu}+\bm{\Gamma}''^A_{\mu\nu}\right) = \frac{\bm{l}_{\mu\nu}}{i\hbar}}.\label{eq:erf_munu}
\end{align}

With this background in mind, the goal of this work is to show how to construct such ETF ($\bGamma'$) and ERF ($\bGamma''$) operators. While the study of ETFs is well explored by now, the case of ERFs is quite unexplored, and we will identify it as a new target below.  The end result of this work will be a compact expression that is easy to implement (Eqs. \ref{eq:center_munu}-\ref{eq:gamma_v1_final},\ref{eq:zeta}), which can easily be added to the rescaling direction in the future so as to maintain the linear and angular momentum of the nuclei during a FSSH calculation.

\section{Theory: Electron Translation Factors (ETFs) and Electron Rotation Factors (ERFs)}
As stated above, the theory of ETFs is well flushed out in the literature, while the concept of ERFs is far less understood.
In order to be as pedagogical as possible, we will now recapitulate the usual prescription for constructing ETFs (whereby one performs an electronic structure calculation in a translating  frame) and then discuss how one might extend these ideas to construct ERFs (whereby one performs an electronic structure calculation in a rotating  frame). More specifically, an outline of this section is as follows: In Sec. \ref{sec:gamma1}, we review the well studied one-electron ETF term ($\bm{\Gamma}'$)  (see Eq. \ref{eq:etf} below). In Sec. \ref{sec:gamma2_general}, we explore the consequences of Eqs. 
\ref{eq:etf_munu} and \ref{eq:erf_munu} (which are constraints on the total $\bm{\Gamma} = \bm{\Gamma}'+\bm{\Gamma}''$), and this exploration leads us to the relevant constraints on $\bm{\Gamma}''$ (see Eqs. \ref{eq:gamma_cond1} and \ref{eq:gamma_cond3} below). 
 While Sec. \ref{sec:gamma2_old} reviews our initial approach\cite{Athavale2023} for constructing one version 
$\bm{\Gamma}''$ (which is found to be unstable), 
Sec. \ref{sec:gamma2_new} offers a new and far more stable ansatz. In Sec. \ref{sec:locality}, we further investigate these new $\bm{\Gamma''}$ matrix elements and show that one can achieve size-consistency by demanding locality of the ERF, which leads to the final expressions for $\bm{\Gamma}''$ shown in Eqs. \ref{eq:center_munu}--\ref{eq:gamma_v1_final}. In Sec. \ref{sec:gamma_alternative}, we briefly demonstrate that the expression we find for $\bm{\Gamma}''$ is not entirely {\em ad hoc} but rather can be derived from a general constrained minimization procedure (as show in Appendix \ref{appendix:lagrangian}). Finally, the special case of the linear molecule is discussed in Sec. \ref{sec:linear}.

\subsection{Translation: $\Gamma'$}\label{sec:gamma1}
In the case of translation, the motivation behind ETFs is to perform electronic structure calculations in a translating basis which leads to so-called ETFs (henceforwared labeled $\bm{\Gamma}'$). As shown in several papers\cite{bates:1958,Schneiderman:1969,Thorson:1978,Delos:1981,Errea:1994,Deumens:1994,Illescas:1998,fatehi:2011:dercouple}, if one boosts all atomic orbitals by the velocity of their attached nucleus, e.g.  
\begin{align}
    \mu(\bx) \rightarrow \mu(\bx) \exp(i\bv_B \cdot \bx/\hbar) 
\end{align}
for an orbital $\mu$ on atom B,
one finds a correction to the derivative couplings of the form: 
\begin{align}
\Gamma'^{A \alpha}_{\mu \nu}  = \frac{1}{2i\hbar} p^{\alpha}_{\mu \nu} \left( \delta_{BA} + \delta_{CA}\right)\label{eq:etf}.
\end{align}
Here and below, \ul{$\mu$ indexes an orbital centered  on atom $B$}, \ul{$\nu$ indexes an  orbital centered on atom $C$}, and $p^{\alpha}_{\mu \nu}$ is the $\alpha$-component of the electronic momentum. Intuitively, the electronic momentum operator emerges because we must take into account the fact that any nuclear displacement moves the electrons as well (as highlighted in Eq. \ref{eq:PJK})\cite{Nafie:1983,ohrn:1994:end,Patchkovskii:2012}. It is easy to show from Eq. \ref{eq:etf} that
\begin{align}
    \sum_A \bm{\Gamma}'^A_{\mu\nu} = \frac{\bm{p}_{\mu\nu}}{i\hbar}.\label{eq:etf_linear}
\end{align}
As far as angular momentum is considered (i.e., Eq. \ref{eq:erf_munu}),  Eq. \ref{eq:etf} implies that
\begin{align}
    \sum_{A\beta\gamma}\epsilon_{\alpha\beta\gamma}X_{A\beta}\Gamma'^{A\gamma}_{\mu\nu} &=\frac{1}{i\hbar}\sum_{\mu\nu}\sum_{\beta\gamma}\braket{\mu}{\frac{1}{2}\epsilon_{\alpha\beta\gamma}\left(X_{B\beta}+X_{C\beta}\right)\hat{p}^\gamma}{\nu}\label{eq:etf_remaining},
\end{align}
where we now have used the Levi-Civita symbol, $\epsilon_{\alpha\beta\gamma}$. 

\subsection{Rotation: $\Gamma''$}\label{sec:gamma2_general}
Beyond translation, the much bigger question regards the proper means to restore angular momentum conservation with $\bm{\Gamma}''$. Given Eq. \ref{eq:etf_linear} and the fact that $\bm{\Gamma} = \bm{\Gamma}' + \bm{\Gamma}''$, Eq. \ref{eq:etf_munu} requires that $\bm{\Gamma}''$ must satisfy
\begin{align}
     \boxed{\sum_A \bm{\Gamma}''^{A}_{\mu\nu} = \bm{0}}.\label{eq:gamma_cond1}
\end{align}

Next, according to Eqs. \ref{eq:erf_munu} and \ref{eq:etf_remaining}, it follows that $\bm{\Gamma}''$ must satisfy
\begin{align}
    \sum_{A\beta\gamma}\epsilon_{\alpha\beta\gamma}X_{A\beta}\Gamma''^{A\gamma}_{\mu\nu} &=\frac{1}{i\hbar}\braket{\mu}{\frac{1}{2}\left(\hat{l}_\alpha^{(B)}+\hat{l}_\alpha^{(C)}\right)}{\nu}.\label{eq:gamma_with_D}
\end{align}
 Here, $\hat{l}_\alpha^{(B)}$ and $\hat{l}_\alpha^{(C)}$ are the $\alpha$-components of the electron angular momentum operators around atoms $B$ and $C$, respectively:
\begin{align}
    \hat{\bm{l}}^{(B)} &= \left(\hat{\bm{x}} - \bm{X}_B\right) \times \hat{\bm{p}},\\
    \hat{\bm{l}}^{(C)} &= \left(\hat{\bm{x}} - \bm{X}_C\right) \times \hat{\bm{p}}.
\end{align}
In compact vector form, Eq. \ref{eq:gamma_with_D} reads: 
\begin{align}
    \boxed{\sum_A\bm{X}_{A}\times\bm{\Gamma}''^{A}_{\mu\nu}  = \frac{1}{i\hbar}\braket{\mu}{\frac{1}{2}\left(\hat{\bm{l}}^{(B)}+\hat{\bm{l}}^{(C)}\right)}{\nu} \equiv \bm{J}_{\mu\nu}}.\label{eq:gamma_cond3}
\end{align}
Here, we have defined \ul{an atom-centered electronic angular momentum} $\bm{J}_{\mu\nu}$; we emphasize that $\bm{J}_{\mu\nu} \ne \frac{1}{i\hbar}\braket{\mu}{\hat{\bm{L}}_e}{\nu}$.

Now, in a recent paper\cite{Athavale2023}, we argued that, because one cannot rotate individual basis functions on a single atom without involving other atoms in the course of a rigid rotation,  one could not generate a strictly local {\em one-electron} ERF operator ($\bm{\Gamma}''^A_{\mu\nu}$) directly analogous to the  ETF  operator ($\bm{\Gamma}'^A_{\mu\nu}$) in Eq. \ref{eq:etf}; here, we would define $\bm{\Gamma}''^A_{\mu\nu}$ to be strictly local if $\bm{\Gamma}''^A_{\mu\nu} = 0 $ when neither $\mu$ nor $\nu$ indexes an orbital centered on atom $A$.  To that end, in Ref. \citenum{Athavale2023}, we constructed a {\em many-electron} strictly local ERF operator that rotates atomic orbitals during the course of a rigid rotation. Note further that any direct  projection of a pre-computed derivative coupling (as in Ref. \citenum{truhlar:2020:angular}) can also be considered a many-electron operator in some sense.  Unfortunately, a many-electron ERF is not desirable -- both  because one loses physical meaning but also because one would like to use such an ERF to build a phase space Hamiltonian (see Ref. \citenum{Coraline:2023:erf}).  To that end, in this paper, we will show below that, if one relaxes strict locality in favor of semi-locality, in fact one can  generate a  one-electron ERF operator $\bm{\Gamma}''^A_{\mu\nu}$.

\subsubsection{Review of the Approach in Ref. \cite{Athavale2023}}\label{sec:gamma2_old}
As means of background, imagine a starting geometry $\bm{X}$ (which is a 3 by $N$ matrix with each column representing the Cartesian coordinate of one atom) and a rotational transformation $\hat{\bm{R}}$, which rotates both the nuclei and the electrons by an angle $\bm{\theta}$ (which is a three-dimensional vectors as it includes the axis of rotation as well as the magnitude). If one wishes to perform a calculation in a basis of rotating electronic atomic orbitals, the key quantity of interest is the angle by which one must rotate all orbital shells of the electronic basis functions.  
To that end, if we assume an infinitesimally small pure rotation, one can calculate\cite{Athavale2023} the angles $d\theta_\alpha$ from the change in nuclear coordiantes,
\begin{align}
    d\bm{X} &= \exp{\left(-\frac{i}{\hbar}\sum_\alpha  d\theta_\alpha\hat{L}^\alpha \right)}\bm{X} -\bm{X}\\
    &=-\frac{i}{\hbar}\sum_\alpha d\theta_\alpha \hat{L}^\alpha \bm{X} \label{eq:dRdtheta},
\end{align}
Here, $\hat{L}$ is the angular momentum operator with matrix elements $\bra{\beta}\hat{L}^\alpha \ket{\gamma} = i\hbar\epsilon_{\alpha\beta\gamma}$ in $\mathbb{R}^3$. 

Now, in the vicinity of a given geometric configuration $\bX$, one separate the geometries that are strict rotations of $\bX$ from the geometries that involve moving interior coordinates.  If one seeks a general angle $\bm{\theta}(\bm{X})$ that is defined for geometries that are not strict rotations of the original configuration $\bX$, the result is not unique. In Ref. \citenum{Athavale2023}, we found an approximate $\bm{\theta}$ by projecting the $3N-$ dimensional problem into a weighted three-dimensional problem, and the final result was: 
\begin{align}
    \pp{\theta_\beta}{X_{A\alpha}} = -\frac{1}{2}\sum_{\gamma\sigma}\epsilon_{\alpha\beta\gamma}\Lambda^{-1}_{\gamma\sigma}X_{A\sigma}
\end{align}
where
\begin{align}
    \Lambda_{\alpha\beta} = \sum_B X_{B\alpha}X_{B\beta}\label{eq:lambda}.
\end{align}
Following the logic in Ref. \citenum{Athavale2023}, this finding would lead us to define a one-electron ERF term as
\begin{align}
    \Gamma''^{A\alpha}_{\mu\nu} &= -\sum_{\beta}\pp{\theta_\beta}{X_{A\alpha}}J_{\mu\nu}^{\beta}\label{eq:gamma_def}\\
    &= \frac{1}{2}\sum_{\beta\gamma\sigma}\epsilon_{\alpha\beta\gamma}J_{\mu\nu}^{\beta}\Lambda^{-1}_{\gamma\sigma}X_{A\sigma} \label{eq:gamma_v1},
\end{align}
For the definition of $\bGamma''$ in Eq. \ref{eq:gamma_v1}, the constraint in Eq. \ref{eq:gamma_cond3} is automatically satisfied.

\subsubsection{An Improved Approach}\label{sec:gamma2_new}
Unfortunately, one can show that the expression $\sum_{\sigma}\Lambda^{-1}_{\gamma\sigma}X_{A\sigma}$  in Eq. \ref{eq:gamma_v1} is very unstable when the atoms are nearly co-planar. While the instability for a {\em linear} molecular might be expected (and be physically meaningful), the instability for a {\em planar} molecule suggests some defects in the expression. To address this problem, here we propose another way of solving Eq. \ref{eq:dRdtheta}. Note that there are 3 by N variables ($d\bm{X}_A$) but only three angles ($\theta_\alpha$). Thus, a  least-squares fit solution would appear to be a strong path forward.  Let us define
\begin{align}
    \tilde{\bm{X}}^\alpha = -\frac{i}{\hbar}\hat{L}^\alpha \bm{X}
\end{align}
and let us solve for $d\theta_\alpha$ by minimizing the squared norm:
\begin{align}
    \left|d\bm{X} - \sum_\alpha \tilde{\bm{X}}^\alpha d\theta_\alpha\right|^2
\end{align}
The solution to this problem is
\begin{align}
    d\theta_\alpha = -\sum_{\beta}K^{-1}_{\alpha\beta}\sum_{A\gamma}\tilde{X}_{A\gamma}^\beta dX_{A\gamma}\label{eq:dtheta_line}
\end{align}
where
\begin{align}
    K_{\alpha\beta} &= -{\rm Tr}\left(\tilde{\bm{X}}^\alpha\tilde{\bm{X}}^{\beta\top}\right)
\end{align}
In differential form, Eq. \ref{eq:dtheta_line} reads:
\begin{align}
    \pp{\theta_\alpha}{X_{A\gamma}} = -\sum_{\beta}K^{-1}_{\alpha\beta}\tilde{X}_{A\gamma}^\beta
\end{align}
Since ${L}^\alpha_{\beta\gamma}=i\hbar \epsilon_{\alpha\beta\gamma}$, the results can then be further simplified:
\begin{align}
    K_{\alpha\beta}   &=-\sum_{A\gamma}X_{A\gamma}X_{A\gamma}\delta_{\alpha\beta} + \sum_{A}X_{A\alpha}X_{A\beta}\label{eq:K}\\
    \pp{\theta_\alpha}{X_{A\gamma}} &= -\sum_{\sigma\beta}K^{-1}_{\alpha\beta}\epsilon_{\sigma\beta\gamma}X_{A\sigma}\label{eq:dtheta_dR_v2}
\end{align}
Substituting Eq. \ref{eq:dtheta_dR_v2} into Eq. \ref{eq:gamma_def}, and noting that $K_{\alpha\beta} = K_{\beta\alpha}$, one recovers
\begin{align}
    \Gamma''^{A\gamma}_{\mu\nu} = \sum_{\alpha\beta\sigma}\epsilon_{\gamma\sigma\beta}X_{A\sigma}K^{-1}_{\beta\alpha}J_{\mu\nu}^{\alpha} \label{eq:gamma_v2}
\end{align}
 The matrix $\bK$ in Eq. \ref{eq:K} is effectively the negative of a massless moment of inertia and can be written in a simple compact vector form:
\begin{align}
    \bm{K} &= -\sum_{A}\left(\bm{X}_A^\top\bm{X}_A\right)\mathcal{I} + \sum_A\bm{X}_A\bm{X}_A^\top\label{eq:K_v2}
\end{align}
where $\bm{X}_A$ is a column vector representing the Cartesian coordinates of atom $A$ and $\mathcal{I}$ is a 3 by 3 identity matrix. The tensor
 $\bGamma''$ in Eq.  \ref{eq:gamma_v2}  also has a simple compact form:
 \begin{align}
    \bm{\Gamma}''^{A}_{\mu\nu} &= \bm{X}_A \times \left(\bm{K}^{-1}\bm{J}_{\mu\nu}\right)\label{eq:gamma_v2_mat}
\end{align}
%  Using the definition of $\bm{\Gamma}''^{A}_{\mu\nu}$ in Eq. \ref{eq:gamma_v2_mat}, one can verify that
% \begin{align}
%     \sum_A \bm{X}_A \times \bm{\Gamma}''^{A}_{\mu\nu} = \bm{J}_{\mu\nu}
% \end{align}
which clearly satisfies the constraint in Eq. \ref{eq:gamma_cond3}.

\subsection{Locality and Size Consistency}\label{sec:locality}
At this point, we have shown how to satisfy Eq. \ref{eq:gamma_cond3}, but we have not addressed the constraint in Eq. \ref{eq:gamma_cond1}. That being said,  before we address such a constraint,  we must first discuss the question of locality.  In particular,  the ansatz for $\bGamma''$ in Eq. \ref{eq:gamma_v2_mat} is incredibly delocalized and not size-consistent.  Physically, if we have two non-interacting subsystems separated far apart from each other, then if atom $A$ resides on one subsystem while orbitals $\chi_\mu$ and $\chi_\nu$ reside on the other subsystem, we will find that $\bm{\Gamma}''^{A}_{\mu\nu} \ne 0$ -- which is unphysical.  To have any physical meaning, $\bm{\Gamma}''^{A}_{\mu\nu}$ must be localized around the atoms where $\chi_\mu,\chi_\nu$ are centered. To achieve a measure of locality, we can introduce a weighting factor $\zeta_{\mu\nu}^A$ such that
\begin{align}
    \bm{K} &\rightarrow \bm{K}_{\mu\nu}=-\sum_{A}\zeta_{\mu\nu}^A\left(\bm{X}_A^\top\bm{X}_A\right)\mathcal{I} + \sum_A\zeta_{\mu\nu}^A\bm{X}_A\bm{X}_A^\top\label{eq:K_scale}\\
    \bm{\Gamma}''^{A}_{\mu\nu} &\rightarrow \zeta_{\mu\nu}^A\bm{X}_A \times \left(\bm{K}_{\mu\nu}^{-1}\bm{J}_{\mu\nu}\right)\label{eq:gamma_v2_scale}
\end{align}
where $\zeta_{\mu\nu}^A$ is maximized when $\chi_\mu$ or $\chi_\nu$ are centered on atom A and decays rapidly otherwise. Eqs. \ref{eq:K_scale}-\ref{eq:gamma_v2_scale} are almost our desired equation for $\bGamma''$, but we have not yet addressed the constraint in Eq. \ref{eq:gamma_cond1}.

In order to satisfy the constraint in Eq. \ref{eq:gamma_cond1}, we will need to recenter the position $\bm{X}_A$  by a quantity $\bm{X}_{\mu\nu}^0$ for each pair of orbitals,  $\chi_\mu$ and $\chi_\nu$.  According to Eq. \ref{eq:gamma_cond1}, we require:
\begin{align}
    \sum_A \zeta_{\mu\nu}^A\left(\bm{X}_A-\bm{X}_{\mu\nu}^0\right) = \bm{0}\label{eq:cond1_scale}
\end{align}
which gives
\begin{align}
    \bm{X}_{\mu\nu}^0 = \sum_A \zeta_{\mu\nu}^A\bm{X}_A /\sum_A \zeta_{\mu\nu}^A.\label{eq:center_munu}
\end{align}
Thus, at the end of the day, a reasonable choice for  $\bm{K}_{\mu\nu}$ and $\bm{\Gamma}''^A_{\mu\nu}$ is:
\begin{align}
    \bm{K}_{\mu\nu}&=-\sum_{A}\zeta_{\mu\nu}^A\left(\bm{X}_A-\bm{X}_{\mu\nu}^0\right)^\top\left(\bm{X}_A-\bm{X}_{\mu\nu}^0\right)\mathcal{I} + \sum_A\zeta_{\mu\nu}^A\left(\bm{X}_A-\bm{X}_{\mu\nu}^0\right)\left(\bm{X}_A-\bm{X}_{\mu\nu}^0\right)^\top\label{eq:final_K}\\
    \bm{\Gamma}''^{A}_{\mu\nu} &= \zeta_{\mu\nu}^A\left(\bm{X}_A-\bm{X}_{\mu\nu}^0\right) \times \left(\bm{K}_{\mu\nu}^{-1}\bm{J}_{\mu\nu}\right),\label{eq:gamma_v1_final}
\end{align}
respectively. Eqs. \ref{eq:center_munu}-\ref{eq:gamma_v1_final} are our final equations for a semi-local one-electron ERF, from which   one can verify that $\bm{\Gamma}''^A_{\mu\nu}$ satisfies Eq. \ref{eq:gamma_cond1} and Eq. \ref{eq:gamma_cond3}:
\begin{align}
    \sum_A \bm{\Gamma}''^{A}_{\mu\nu} &= \sum_A \zeta_{\mu\nu}^A\left(\bm{X}_A-\bm{X}_{\mu\nu}^0\right) \times \left(\bm{K}_{\mu\nu}^{-1}\bm{J}_{\mu\nu}\right) = 0\\    
    \sum_A \bm{X}_A \times \bm{\Gamma}''^{A}_{\mu\nu} &=\sum_A \zeta_{\mu\nu}^A\bm{X}_A \times \left[\left(\bm{X}_A-\bm{X}_{\mu\nu}^0\right) \times \left(\bm{K}_{\mu\nu}^{-1}\bm{J}_{\mu\nu}\right)\right]\\
    &=\sum_A \zeta_{\mu\nu}^A\left(\bm{X}_A-\bm{X}_{\mu\nu}^0\right) \times \left[\left(\bm{X}_A-\bm{X}_{\mu\nu}^0\right) \times \left(\bm{K}_{\mu\nu}^{-1}\bm{J}_{\mu\nu}\right)\right]\\
    &=\bm{J}_{\mu\nu}
\end{align}

\subsubsection{The Choice of $\zeta^A_{\mu\nu}$}
All that remains is to choose a function form for $\zeta^A_{\mu\nu}$ in  Eqs. \ref{eq:center_munu}--\ref{eq:gamma_v1_final}. Below, we investigate a semi-local function of the form:  
\begin{align}
    \zeta^A_{\mu\nu} &= \exp\left(-w \frac{2|(\bm{X}_A-\bm{X}_B)|^2 |(\bm{X}_A-\bm{X}_C)|^2}{|(\bm{X}_A-\bm{X}_B)|^2 + |(\bm{X}_A-\bm{X}_C)|^2}\right)\label{eq:zeta}
\end{align}
where again we assume $\chi_\mu$ is centered on atom $B$ and $\chi_\nu$ is centered on atom $C$. The parameter $w$ controls the locality of the final ERF, and below we will provide insight into how to best optimize and analyze such a function. 

\subsubsection{An Alternative Approach Based on Minimization}\label{sec:gamma_alternative}
Interestingly, Eqs. \ref{eq:center_munu}--\ref{eq:gamma_v1_final} above for $\bGamma''^A_{\mu \nu}$ can be derived from a totally different principle in a more direct fashion. 
The idea is to compute the minimal $\bm{\Gamma}''^A_{\mu\nu}$ that are consistent with the constraints in Eqs. \ref{eq:gamma_cond1} and \ref{eq:gamma_cond3}. The corresponding Lagrangian is:
\begin{align}
    &\mathcal{L} = \sum_{A \mu \nu} \frac{1}{\zeta^{A}_{\mu\nu}}\bm{\Gamma}''^{A \top}_{\mu \nu}\bm{\Gamma}''^{A}_{\mu \nu} - \sum_{\mu \nu}\bm{\lambda}_{1\mu\nu}^{\top}\left(\sum_A\bm{\Gamma}''^{A}_{\mu \nu}\right) -\sum_{\mu \nu}\bm{\lambda}_{2\mu\nu}^{\top} \left(\sum_{A} \bm{X}_A \times \bm{\Gamma}''^{A}_{\mu \nu}-\bm{J}_{\mu \nu}\right)\label{eq:lagrangian}
\end{align}
Here,  $\zeta^{A}_{\mu\nu}$ is the weighting factor and the constraints controlled by $\bm{\lambda}_{1\mu\nu}$ and $\bm{\lambda}_{2\mu\nu}$ are Eqs. \ref{eq:gamma_cond1} and \ref{eq:gamma_cond3}, respectively. As shown in Appendix \ref{appendix:lagrangian},  minimizing the Lagrangian in Eq. \ref{eq:lagrangian} is identical to Eqs. \ref{eq:center_munu}--\ref{eq:gamma_v1_final}.

\subsection{Case of linear molecule}\label{sec:linear}
Before providing numerical results, one special case must be addressed, for which Eq. \ref{eq:gamma_v1_final} needs to be revised: namely, the case of a linear molecule. In such a case, a rotation around the molecular axis is redundant which leads to  troubles for the form the ERF calculated  in Eq. \ref{eq:gamma_v1_final}. Specifically, $\bm{K}_{\mu\nu}$ is not invertible. We can address this issue by assuming that the ERF term should  recover $\bm{J}_{\mu\nu}$ only in the directions perpendicular to the molecular axis. After all, rotating the nuclei along the molecular axis does not change the electron angular momentum.

Mathematically, this assumption allows us to exclude the null-space of $\bm{K}_{\mu\nu}$ when calculating $\bm{K}_{\mu\nu}^{-1}\bm{J}_{\mu\nu}$ in Eq. \ref{eq:gamma_v1_final}. Specifically, Let $\bm{u}_1,\bm{u}_2,\bm{u}_3$ be the complete orthonormal basis of $\mathbb{R}^3$ and $\bm{u}_3$ is along the molecular axis. Since $\bm{X}_A-\bm{X}_{\mu\nu}^0$ is parallel to $\bm{u}_3$, we may write
\begin{align}
    \bm{X}_A-\bm{X}_{\mu\nu}^0 = x_{\mu\nu}^A \bm{u}_3
\end{align}
and 
\begin{align}
    \bm{K}_{\mu\nu}&=-\sum_{A}\zeta_{\mu\nu}^A\left(\bm{X}_A-\bm{X}_{\mu\nu}^0\right)^\top\left(\bm{X}_A-\bm{X}_{\mu\nu}^0\right)\mathcal{I} + \sum_A\zeta_{\mu\nu}^A\left(\bm{X}_A-\bm{X}_{\mu\nu}^0\right)\left(\bm{X}_A-\bm{X}_{\mu\nu}^0\right)^\top\\
    &=-\sum_A\zeta_{\mu\nu}^A\left(x_{\mu\nu}^{A}\right)^2\mathcal{I}+\sum_A\zeta_{\mu\nu}^A\left(x_{\mu\nu}^{A}\right)^2\bm{u}_3\bm{u}_3^\top\\
    &=-\sum_A\zeta_{\mu\nu}^A\left(x_{\mu\nu}^{A}\right)^2\left(\mathcal{I}-\bm{u}_3\bm{u}_3^\top\right)
\end{align}
Clearly, $\bm{u}_1$ and $\bm{u}_2$ are the two degenerate eigenvectors of $\bm{K}_{\mu\nu}$ with the eigenvalue $-\sum_A\zeta_{\mu\nu}^A\left(x_{\mu\nu}^{A}\right)^2$, while $\bm{u}_3$ has the corresponding eigenvalue of zero. Consequently, for a linear molecule, we simply replace $\bm{K}_{\mu\nu}^{-1}\bm{J}_{\mu\nu}$ in Eq. \ref{eq:gamma_v1_final} with
\begin{align}
    \bm{K}_{\mu\nu}^{-1}\bm{J}_{\mu\nu} \rightarrow -\left(\sum_A\zeta_{\mu\nu}^A\left(x_{\mu\nu}^{A}\right)^2\right)^{-1}\left(\mathcal{I}-\bm{u}_3\bm{u}_3^\top\right)\bm{J}_{\mu\nu}
\end{align}
In our developmental version of the Q-Chem electronic structure package\cite{Epifanovsky2021}, we have implemented two different pieces of code: one which for the polyatomic case and one for the linear case. Presumably, if an advanced solver with a generalized inversion routine were available that can solve $Ax=b$ for A not invertible, both cases can be combined into one code.

\section{Numerical Results and Discussions}

\begin{figure*} [ht]
\centering
\includegraphics[width=0.5\textwidth]{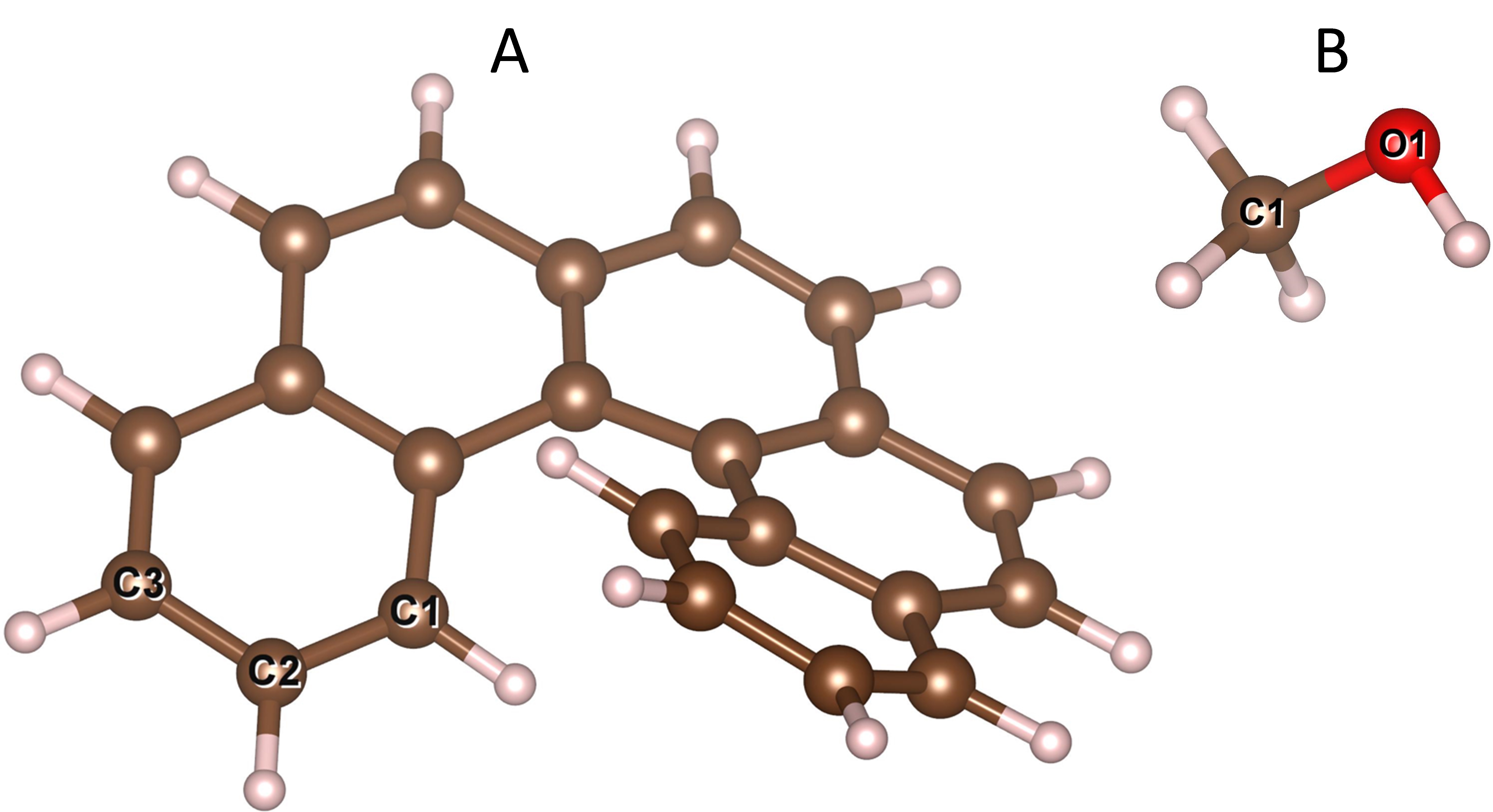}
\caption{Systems that the ERF term $\bm{\Gamma}''$ is calculated for. (A) [5]Helicene. (B) Methanol. The Cartesian coordinates are provided in Appendix \ref{appendix:geometry}.}
\label{fig:geometry}
\end{figure*}

The choice of $w$ is critical for determining a meaningful ERF. On the one hand, $w$ should not be too small; $w$ controls the locality of the ERF term and setting $w$ to zero will lead to complete delocalization (which breaks size consistency). On the other hand, an arbitrarily large value is not desirable either, as such a choice would force many molecular environments to appear as if they were diatomic (which we argued above is unstable and equivalent to enforcing strict locality). 
%Such a choice enforces ERF localize entirely on the two(one) atoms where $\mu$ and $\nu$ are centered and the system becomes practically diatomic (uni-atomic) for each $\mu,\nu$ pair. Since the rotation of a diatomic (uni-atomic) system is not fully determined, and different from the linear molecule case discussed in Section \ref{sec:linear}, the electronic angular momentum along the effective diatomic axis matters, $\bm{\Gamma}''^A_{\mu\nu}$ calculated in this way is deficient. 
From a numerical perspective, an arbitrarily large $w$ will force the $\bm{K}_{\mu\nu}$ matrix to become singular, causing numerical instability and a violation of the constraints in Eqs. \ref{eq:gamma_cond1} and \ref{eq:gamma_cond3}. 

\begin{figure*} [ht]
\centering
\begin{subfigure}[b]{0.495\linewidth}
\centering
\includegraphics[width=1.0\textwidth]{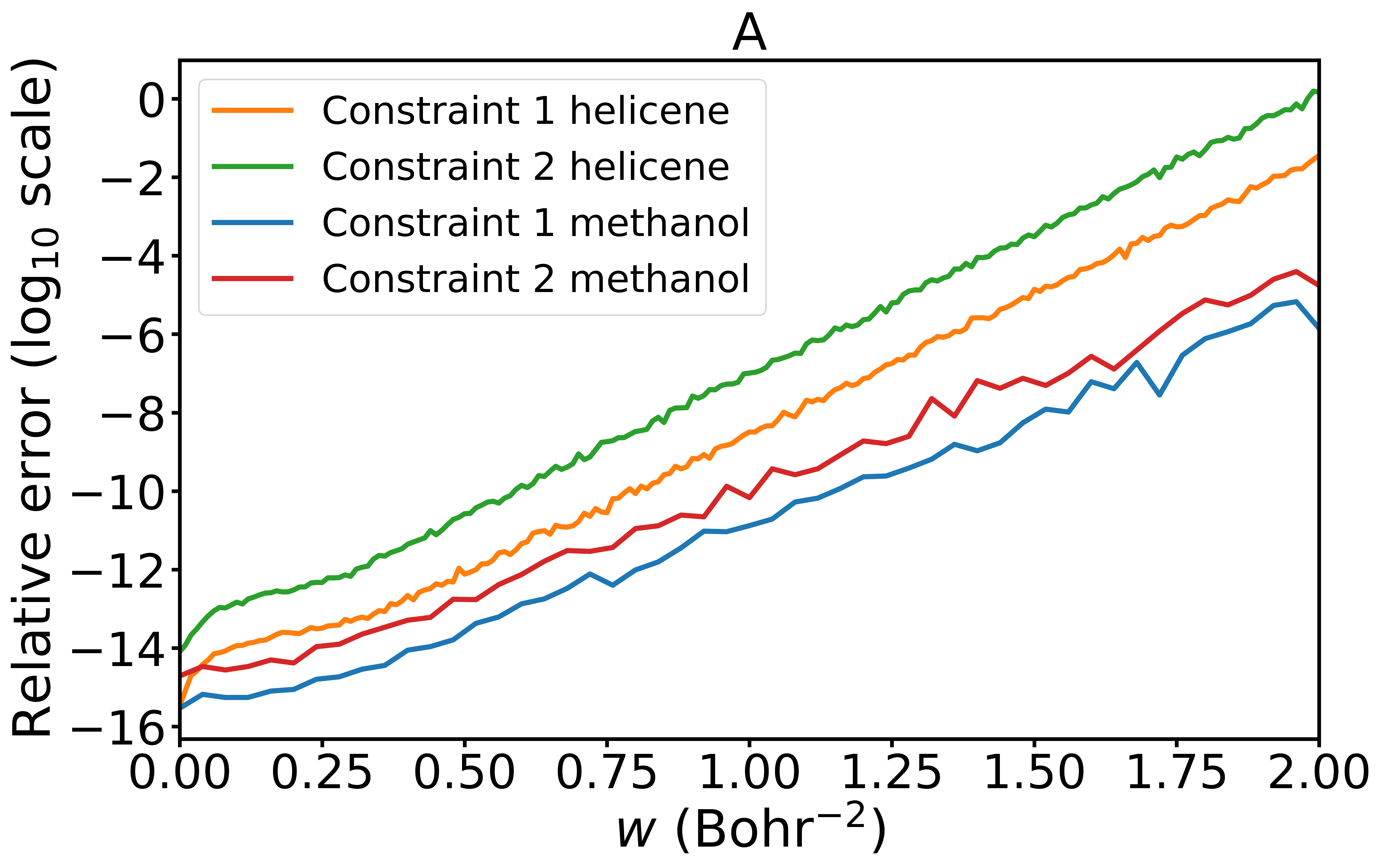}
\end{subfigure}
\begin{subfigure}[b]{0.495\linewidth}
\centering
\includegraphics[width=1.0\textwidth]{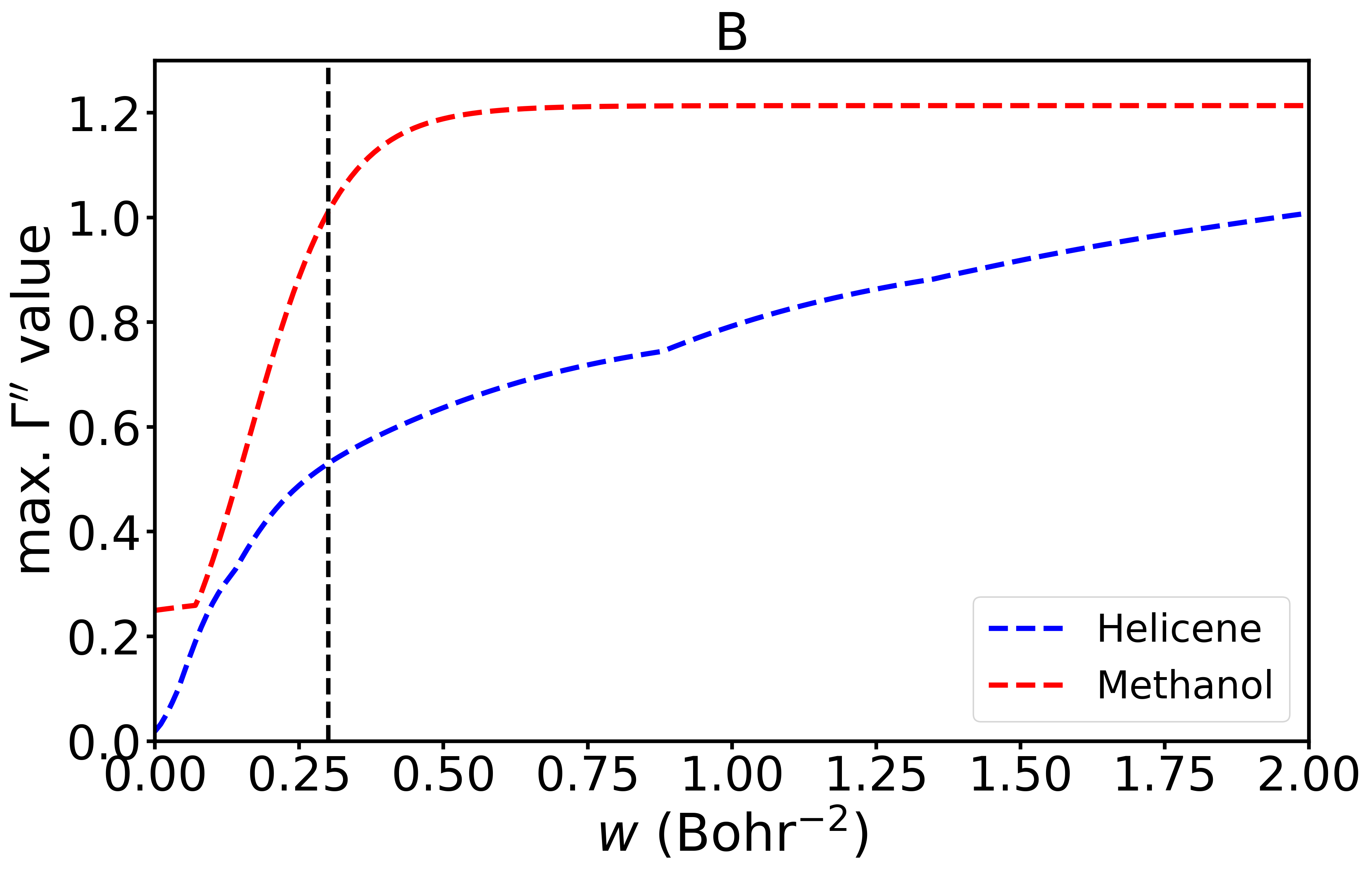}
\end{subfigure}
\begin{subfigure}[b]{0.475\linewidth}
\centering
\includegraphics[width=1.0\textwidth]{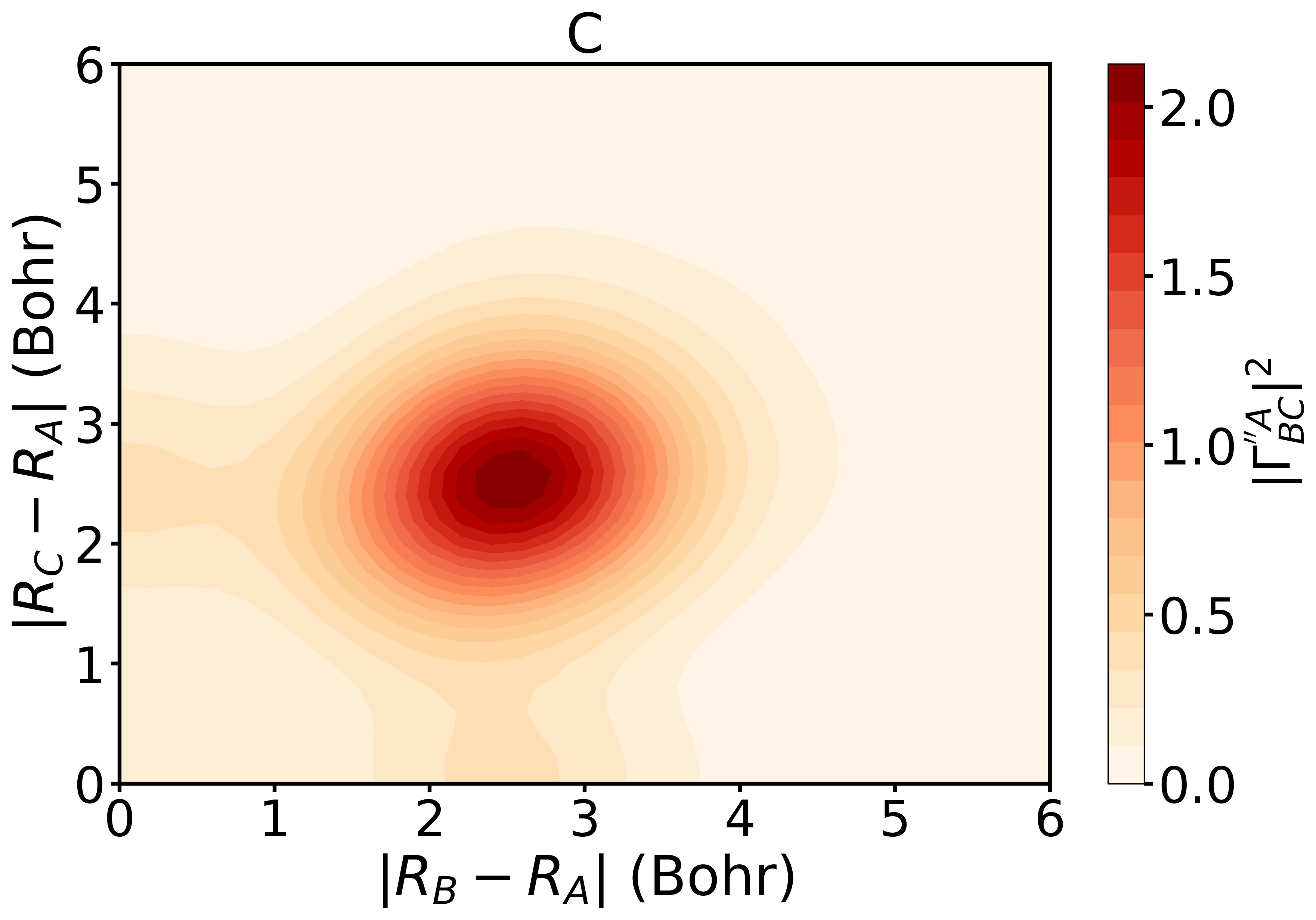}
\end{subfigure}
\begin{subfigure}[b]{0.495\linewidth}
\centering
\includegraphics[width=1.0\textwidth]{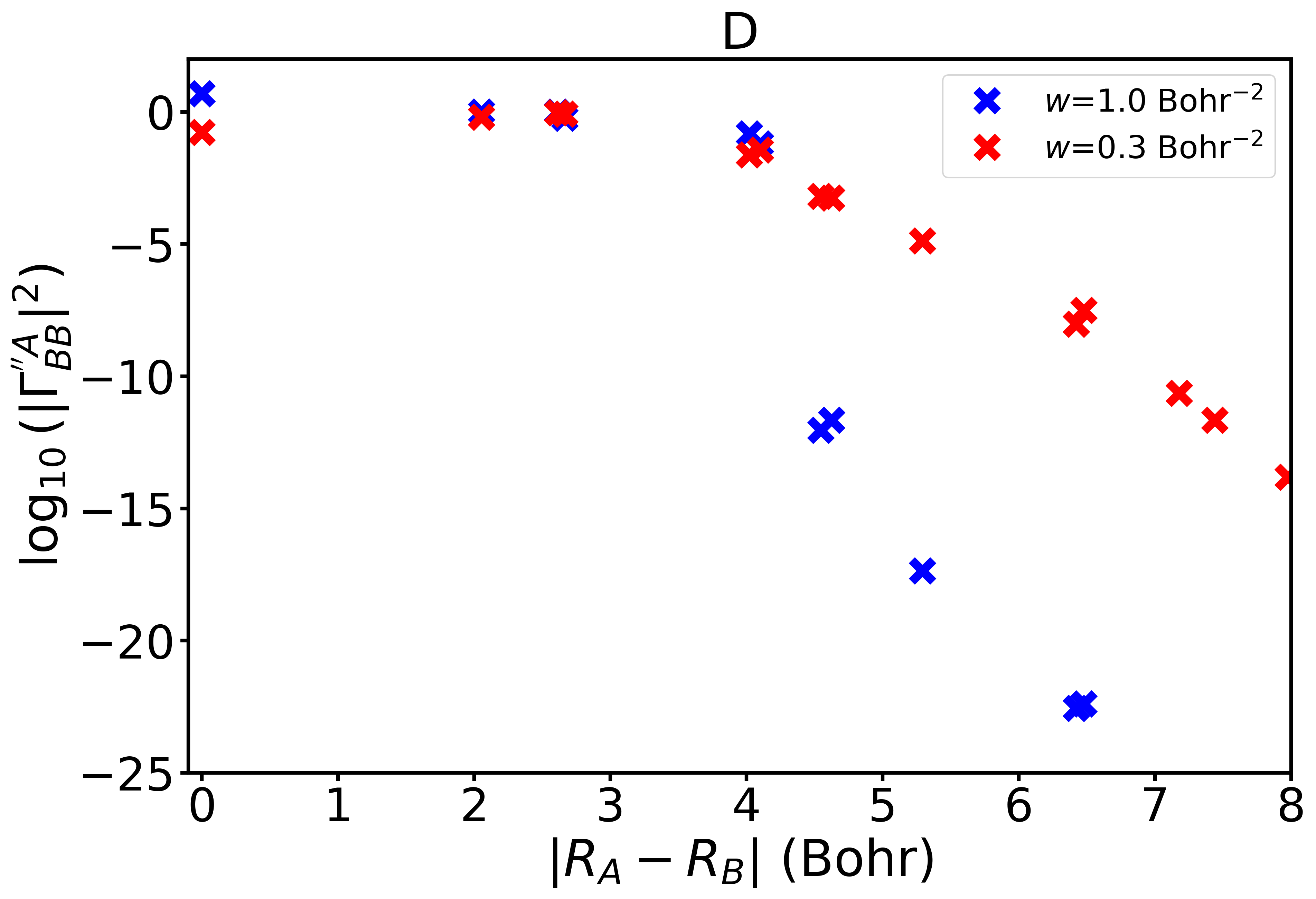}
\end{subfigure}
\caption{Numerical stability and locality of $\bm{\Gamma}''^A_{\mu\nu}$. (A) Errors in the constraint in Eq. \ref{eq:gamma_cond1} (marked as Constraint 1) and Eq. \ref{eq:gamma_cond3} (marked as Constraint 2) as a function of different $w$ values. (B) Maximum value of $\bm{\Gamma}''^{A}_{\mu\nu}$  (in atomic units) as a function of $w$. The dashed vertical line represents $w = 0.3$ ${\rm Bohr}^{-2}$. (C) Heat-map for $\left|\Gamma''^A_{BC}\right|^2$ as defined in Eq. \ref{eq:gamma_BC} with atom $A$ from Eq. \ref{eq:gamma_BC} fixed as C2 in Fig. \ref{fig:geometry}A. A Gaussian broadening with $\sigma^2 = \frac{1}{2}$ ${\rm Bohr}^2$ is applied. (D) Decay of $\left|\Gamma''^A_{BB}\right|^2$ as a function of the distance between atom 
 $A$ and atom $B$.}
\label{fig:locality}
\end{figure*}

To demonstrate this point, we have applied our algorithm to two systems, namely the [5]helicene and methanol molecules (shown in Fig. \ref{fig:geometry}).  In Fig. \ref{fig:locality}A, we plot the errors in the two constraints (Eqs. \ref{eq:gamma_cond1} and \ref{eq:gamma_cond3}) for different $w$ values.  We find that the error in the two constraints increases exponentially as $w$ becomes larger, and the deviation to the constraint in Eq. \ref{eq:gamma_cond1} reaches $10^{-7}$ when $w$ is greater than 1 ${\rm Bohr}^{-2}$. Next, in Fig. \ref{fig:locality}B, we plot the maximum value of $\bm{\Gamma}''$  versus $w$. The maximum value of $\bm{\Gamma}''$ grows rapidly when $w$ changes from 0 to $\approx 0.3$ ${\rm Bohr}^{-2}$ and then slows down. These two characteristics suggests that $w = 0.3$ ${\rm Bohr}^{-2}$ is a safe choice that balances both locality and numerical stability. 
To provide further insights into the locality of the $\bm{\Gamma}''$ tensor, see Fig. \ref{fig:locality}C.  Here, we define a quantity
\begin{align}
\left|\Gamma''^A_{BC}\right|^2 = \sum_{\substack{\rm \mu\ on\ B \\ \rm \nu\ on\ C}}\left|\bm{\Gamma}''^A_{\mu\nu}\right|^2\label{eq:gamma_BC}
\end{align}
and 
visualize $\bGamma''$ in terms of the distances between atom $A$ and atom $B,C$. 
More specifically, we plot  a heat-map that spans over all possible $B,C$ pairs for [5]helicene with atom $A$ fixed as C2 labeled in Fig. \ref{fig:geometry}A. The heat-map plots $\bm{\Gamma}''$ calculated with $w = 0.3$ ${\rm Bohr}^{-2}$; a Gaussian broadening function (with $\sigma^2 = \frac{1}{2}$ ${\rm Bohr}^2$) is applied for smoothness. In Fig. \ref{fig:locality}D, we plot $\left|\Gamma''^A_{BC}\right|^2$ with $B=C$  without the Gaussian broadening function, so as to provide the most precise view possible for the decay of  $\bm{\Gamma}_{\mu\nu}''^A$.

\section{Discussion: Invariance of ${\Gamma}$ under translation and rotation}\label{sec:disscusion}

Before concluding this manuscript, a discussion of translational and rotational invariance is appropriate.  Obviously, in order to apply an ETF or ERF in a meaningful fashion, the matrix elements $\bm{\Gamma}_{JK}$ should not depend on the origin or orientation of the molecule. Unfortunately, establishing such translational and especially rotational invariance is complicated by the fact that atomic orbitals come in shells and does not rotate with the molecular frame. For instance, a $p_x$ atomic orbital in one orientation becomes a $p_y$ atomic orbital when rotating the molecule by $90^\circ$ along the $z$-axis.  Now, quite generally, in any quantum chemistry calculation, all calculations depend on the vector space of atomic orbitals (and not on the individual choice of basis functions), which explains why quantum chemical molecular energies are rotationally invariant. This fact can most easily be seen by noting that $h_{\mu\nu}$ transforms as a well-defined tensor operator, and the creation/annihilation operators $a_{\mu}^{\dagger}$/$a_{\nu}$ transform as vectors. Thus,  the one-electronic Hamiltonian,
\begin{align}
    \sum_{\mu \nu} h_{\mu \nu} a_{\mu}^{\dagger} a_{\nu},
\end{align}
is invariant to basis, i.e.  one can mix one set of atomic orbitals into any other set of basis functions without changing the overall Hamiltonian.   Now, obviously, if one considers the operator 
\begin{align}
    \sum_{\mu \nu} \bGamma^A_{\mu \nu} a_{\mu}^{\dagger} a_{\nu},
\end{align}
mixing basis functions on different atoms does not make much sense -- because the operator itself depend on a given atom A -- but mixing basis functions on the same atom does not change the overall operator.   Thus, one would hope that such a mixing does not affect any momentum-rescaling results. Indeed, in Appendix \ref{appendix:trans} and \ref{appendix:rotation}, we will show that rescaling direction, $\bm{\Gamma}_{JK}$ is indeed invariant to translations and rotations of the molecule.

%or rotation of the molecule.

%we find (as we sought):
%\begin{align}
%\bGamma''_{\mu \nu} \cdot \bP_N = 
%\bGamma''_{\bar{\mu} \bar{\nu}} \cdot \bar{\bP}_N \end{align}

%In other words, $\bm{\Gamma}'' \cdot \bP_N$ is invariant in a rotation basis that rotates with the molecule and therefore the physics of the electronic structure will remain unchanged provided the basis itself is composed of full shells that are rotationally invariant around every atom (which is always the case). In this sense, even though our expressions for $\bm{\Gamma}''$ (i.e., Eqs. \ref{eq:center_munu}-\ref{eq:gamma_v1_final}) is not written in a rotating AO basis, there should be no change in the physical quantity $\bm{\Gamma}''_{JK}\cdot \bP_N = \sum_{\mu\nu}D_{\mu\nu}^{JK}\bm{\Gamma}''_{\mu\nu}\cdot \bP_N$ (where $D_{\mu\nu}^{JK}$ is the density matrix) if we rotate the entire molecule.

\section{Conclusions and Outlook}
By working in a traveling and rotating basis, we have shown that one can derive physically motivated {\em one-electron} ETF ($\bm{\Gamma}'$) and ERF  ($\bm{\Gamma}''$) operators so as to account for electronic motion. 
While the ETF in Eq. \ref{eq:etf} is well-known, the key new equations of this communication are Eqs. \ref{eq:center_munu}-\ref{eq:gamma_v1_final}.  An alternative derivation of the ERF operator (as found by a constrained minimization) is offered in the Appendix as well.  
Perhaps not surprisingly, while the ETFs involve the electronic linear momentum operator, the ERFs involve the electronic angular momentum operator. 

As discussed in Sec. \ref{sec:locality}, although $\bm{\Gamma}'^A_{\mu\nu}$ can be constructed in a strictly local fashion,  the $\bm{\Gamma}''^A_{\mu \nu}$ tensor  can be constructed only in a semilocal fashion. This difference is inevitable given the different nature of linear versus angular momentum, but indeed a reasonably semi-localized (not strictly localized) $\bm{\Gamma}''$ can be achieved by enforcing locality through the $\zeta$ weighting factor in Eq. \ref{eq:zeta}.  As a practical matter, the data in Fig. \ref{fig:locality}B suggests that $w = 0.3  \;{\rm Bohr}^{-2}$ is a reasonable choice. 
Note that the  one electron operator ERFs derived here should be applicable to 
just about any excited states including TD-DFT/TDHF states, where the community has established how to interpret the relevant response functions (at least approximately) through the lense of wavefunctions\cite{ou:2014:tddft_rpa,alguire:2014:tdhf, ou:2015:dc_response_theory, herbert:2014:jcp_dercouple}.
Interestingly, by enforcing locality (or semi-locality) -- which is meaningful as far as achieving size consistency -- the matrix elements of the ERFs increase, such that,  for the molecules presented here, the ERFs between different CIS excited states are roughly the same order of magnitude as the corresponding ETFs (which contrasts with the results in Ref. \cite{Athavale2023}). 

Looking forward, it is important to note that the approach above can be easily extended to  systems with spin degrees of freedom if we remember that electronic spin is an important form of angular momentum. In such a case, if we wish to conserve the total angular momentum, we need only define 
\begin{align}
    J_{\mu\nu}^{\alpha} = \frac{1}{i\hbar}\braket{\mu}{\frac{1}{2}\left(\hat{l}_\alpha^{(B)}+\hat{l}_\alpha^{(C)}\right)+\hat{s}_\alpha}{\nu}
\end{align}
instead of Eq. \ref{eq:gamma_cond3}, where now we work with a spin-atomic basis (instead of a spatial orbital basis) and allow for the ERFs to mix spin degrees of freedom.

Finally, in a companion paper\cite{Coraline:2023:erf}, we  argue that the ERFs and ETFs proposed in the present paper should have a value far beyond the present context of momentum-rescaling in surface hopping.  In particular, as shown in Ref. \citenum{Bian:2023}, one can argue that standard (classical) Born-Oppenheimer dynamics (without a Berry force) ignore electronic dynamics and therefore do not conserve the total angular or linear momentum in general. In such a context, however, Ref. \citenum{Coraline:2023:erf} demonstrates that when dynamics are run along a Hamiltonian parameterized by nuclear position and momentum, $\hat{H}(\bm{X},\bm{P}) = \frac{\bm{P}^2}{2\bm{M}} - i\hbar\frac{\bm{P}}{\bm{M}}\cdot\bm{\Gamma} +\hat{H}_{\rm el}(\bm{X})$, the resulting dynamics {\em do} conserve the total linear and angular momentum.  Thus, the present derivation of $\bm{\Gamma}$ may well be extremely important in the future for adiabatic propagation -- and not just for surface-hopping momentum rescaling. Moreover, Truhlar and co-workers have demonstrated that Ehrenfest dynamics violate angular momentum conservation, and they have suggested removing the relevant term  from the derivative coupling that appears in the Ehrenfest equation of motion.  Thus, the present derivation of $\bm{\Gamma}$ should also important in the future for non-adiabatic propagation more generally  (although we would submit that a better remedy for Ehrenfest dynamics is to include the non-Abelian Berry curvature\cite{Coraline:2024:nonabelian})
.  Looking forward, our hope is that the present ERFs will be useful for modeling coupling nuclear-electronic-spin dynamics quite generally, potentially for modeling the chiral-induced spin selectivity (CISS) effect\cite{naaman:2012:jpcl}.

%Lastly, we believe the most interesting question of all for this discussion is: Is there any physical meaning to the ETFs and ERFS presented here, beyond the realm of surface hopping?  This subject will be taken up in a subsequent paper.
\section{Acknowledgements}
This work is supported by the National Science Foundation under Grant No. CHE-2102402.

\section{Appendix}

\subsection{Translational Invariance}\label{appendix:trans}
As discussed in Sec. \ref{sec:disscusion}, one would like to be sure that within any surface-hopping algorithm, the momentum-rescaling direction does not depend on the orientation or origin of the chemical problem. 
To that end, let us here demonstrate translational invariance.
To begin our discussion, let us emphasize that the one electron Hamiltonian is of course invariant to translation of the molecule. This fact is clear when we recognize that, upon translation, the atomic orbitals translate with the molecule
so that
\begin{align}
    h_{\mu \nu}(\bX + \delta \bX) = h_{\mu \nu}(\bX),
\end{align}
and therefore the density matrix between any two electronic states is also unchanged
\begin{align}
    D^{JK}_{\mu \nu}(\bX + \delta \bX) = D^{JK}_{\mu \nu}(\bX).
\end{align}
Hence, it follows that:
\begin{align}
    \left<J \middle|\sum_{\mu \nu} h_{\mu \nu} a_{\mu}^{\dagger} a_{\nu}  \middle| K \right>\Biggl|_{\bX_0} =
    \left<J \middle|\sum_{\mu \nu} h_{\mu \nu} a_{\mu}^{\dagger} a_{\nu}  \middle| K \right>\Biggl|_{\bX_0+\delta X}
\end{align}

Next, consider rescaling the momentum along the proposed $\bGamma = \bGamma' + \bGamma''$ direction, where the ETF is defined in Eq. \ref{eq:etf} and the ERF is defined in Eqs. \ref{eq:center_munu}-\ref{eq:gamma_v1_final}:
\begin{align}
    \left<J \middle|\sum_{\mu \nu} \bGamma^A_{\mu \nu} a_{\mu}^{\dagger} a_{\nu} \cdot \bP \middle| K \right>
    \label{eq:Gamma_trans}
\end{align}
Note that under translation, the following rules hold:

\begin{enumerate}
    \item  $\bP$ does not change direction. 

\item  $\bm{\Gamma}'$ is translational invariant ($\bGamma'^A_{\mu \nu}(\bX + \delta \bX) = \bGamma'^A_{\mu \nu}(\bX)$) because   $\bp_{\mu \nu}(\bX + \delta \bX) = \bp_{\mu \nu}(\bX)$.  

\item  $\bm{K}_{\mu\nu}$  and $\bm{J}_{\mu \nu}$ are both invariant under translation, i.e. $\bm{K}_{\mu \nu}(\bX + \delta \bX) = \bm{K}_{\mu \nu}(\bX)$ and $\bm{J}_{\mu \nu}(\bX + \delta \bX) = \bm{J}_{\mu \nu}(\bX)$,  
so that $\bm{\Gamma}''^A_{\mu\nu}$ is  also translationally invariant (($\bGamma''^A_{\mu \nu}(\bX + \delta \bX) = \bGamma''^A_{\mu \nu}(\bX)$)).
%because:  $(i)$ for a translation such that $\bm{X}_A \rightarrow \bm{X}_A - \bm{X}^o$, the scaled center for each $\mu,\nu$ is changed to $\bm{X}_{\mu\nu}^0 - \bm{X}^o$ so that
\end{enumerate}

These rules prove that
\begin{align}
    \boxed{\bGamma^A_{\mu \nu}(\bX + \delta \bX) = \bGamma^A_{\mu \nu}(\bX)}.
\end{align}

\subsection{Rotational Invariance}\label{appendix:rotation}

The final item that remains to be proven is rotational invariance. 
Proving rotational invariance is a bit more involved than for translation because, even though a Gaussian basis in a quantum chemistry code translates with the molecule, the basis does \ul{not} rotate with the molecule. In other words, in practice, the orientation of a given atomic orbital does not  depend on the orientation of the molecule. 
To that end, establishing notation will be essential. Let $\chi_\mu$ be an atomic orbital centered on atom $B$ with a definitive orientation, e.g. a $p_x$ orbital. If the molecule translate to a new location, we will still index the same orbital by $\chi_\mu$ (which would still be, e.g., a $p_x$ orbital).  Now, if the molecule rotates, let us denote the rotated atomic orbital by $\bar{\chi_\mu},\bar{\chi_\nu}$.  
Let us represent a rotational transformation of $\bm{X}^A$  by a matrix $\bm{R}$, i.e.,
\begin{align}
    \bm{X}^A &\rightarrow \bm{R}\bm{X}^A
    \label{eq:barX}
\end{align}

To begin our discussion,
consider  the one-electron Hamiltonian, $h_{\mu\nu}$. These matrix elements are rotationally invariant 
\begin{align}
    h_{\bar{\mu}\bar{\nu}} (\bR \bX) &= h_{\mu\nu} (\bX),
\end{align}
which forces the corresponding transition density matrix to also be  invariant
\begin{align}
    D^{JK}_{\bar{\mu}\bar{\nu}} (\bR \bX) &= D^{JK}_{\mu\nu} (\bX).\label{eq:Drot}
\end{align}
Eq. \ref{eq:Drot} reflects the fact that the states $J$ and $K$ rotate with the molecule and the same electronic structure solutions must arise at any geometry in the presence of identical Hamiltonian matrix elements. Altogether, it then follows that:
\begin{align}
    \left<J \middle|\sum_{\mu \nu} h_{\bar{\mu} \bar{\nu}} a_{\bar{\mu}}^{\dagger} a_{\bar{\nu}}  \middle| K \right>\Biggl|_{\bR \bX} =
    \left<J \middle|\sum_{\mu \nu} h_{\mu \nu} a_{\mu}^{\dagger} a_{\nu}  \middle| K \right>\Biggl|_{\bX}
\end{align}

Next, let us consider the proposed one-electron ETF and ERF terms. We would like to show that these tensors  lead to rotationally invariant directions in the sense that:
\begin{align}  
    \label{eq:final1}
    \left<J \middle|\sum_{\mu \nu} \bGamma'^A_{\bar{\mu} \bar{\nu}} a_{\bar{\mu}}^{\dagger} a_{\bar{\nu}} 
    \cdot \bP \middle| K \right>\Biggl|_{\bR \bX} &= 
    \left<J \middle|\sum_{\mu \nu} \bGamma'^A_{\mu \nu} a_{\mu}^{\dagger} a_{\nu} \cdot \bP \middle| K \right>\Biggl|_{\bX} \\
       \label{eq:final2}
       \left<J \middle|\sum_{\mu \nu} \bGamma''^A_{\bar{\mu} \bar{\nu}} a_{\bar{\mu}}^{\dagger} a_{\bar{\nu}} 
       \cdot \bP \middle| K \right>\Biggl|_{\bR \bX} &= 
    \left<J \middle|\sum_{\mu \nu} \bGamma''^A_{\mu \nu} a_{\mu}^{\dagger} a_{\nu} \cdot \bP \middle| K \right>\Biggl|_{\bX}
\end{align}

To that end, note that if we rotate a molecule, it must be true that
\begin{align}
    \bm{P}^A &\rightarrow \bm{R}\bm{P}^A
    \label{eq:barP},
\end{align}
and it is also straightforward to show that
    \begin{align}
    \label{eq:barp}
\bm{p}_{\bar{\mu}\bar{\nu}} &= \bm{R}\bm{p}_{\mu\nu} \\
    \label{eq:barJ}
\bm{J}_{\bar{\mu}\bar{\nu}} &= \bm{R}\bm{J}_{\mu\nu}
\end{align}
This equality is also proved explicitly in the Appendix of Ref. \citenum{Coraline:2023:erf}.
At this point, Eq. \ref{eq:final1} follows from the definition in Eq. \ref{eq:etf} and the rotational transformations in Eqs. \ref{eq:Drot}, \ref{eq:barP}, and \ref{eq:barp}.

Furthermore, from Eqs. \ref{eq:center_munu}  and \ref{eq:final_K}, it follows that:
\begin{align}
    \bm{X}_{\bar{\mu}\bar{\nu}}^0 &= \sum_A \zeta_{\mu\nu}^A\bm{R}\bm{X}_A /\sum_A \zeta_{\mu\nu}^A\\
    &=\bm{R}\bm{X}_{\mu\nu}^0\\
    \bm{K}_{\bar{\mu}\bar{\nu}}&=-\sum_{A}\zeta_{\mu\nu}^A\left(\bm{X}_A-\bm{X}_{\mu\nu}^0\right)^\top\left(\bm{X}_A-\bm{X}_{\mu\nu}^0\right)\mathcal{I} \nonumber\\
    &\ \ \ \ + \sum_A\zeta_{\mu\nu}^A\bm{R}\left(\bm{X}_A-\bm{X}_{\mu\nu}^0\right)\left(\bm{X}_A-\bm{X}_{\mu\nu}^0\right)^\top\bm{R}^\top\\
    &=\bm{R}\bm{K}_{\mu\nu}\bm{R}^\top
    \end{align}

Substituting the above equations into Eq. \ref{eq:gamma_v1_final}, we find 
\begin{align} \bm{\Gamma}''^A_{\bar{\mu}\bar{\nu}} &= \zeta_{\mu\nu}^A\left(\bm{R}\left(\bm{X}_A-\bm{X}_{\mu\nu}^0\right)\right) \times \left(\bm{R}\bm{K}_{\mu\nu}^{-1}\bm{R}^\top\bm{R}\bm{J}_{\mu\nu}\right)\\
    &=\zeta_{\mu\nu}^A\left(\bm{R}\left(\bm{X}_A-\bm{X}_{\mu\nu}^0\right)\right) \times \left(\bm{R}\bm{K}_{\mu\nu}^{-1}\bm{J}_{\mu\nu}\right)\\
    &=\zeta_{\mu\nu}^A\bm{R}\left(\left(\bm{X}_A-\bm{X}_{\mu\nu}^0\right)\times \left(\bm{K}_{\mu\nu}^{-1}\bm{J}_{\mu\nu}\right)\right)\\
    &=\bm{R}\bm{\Gamma}''^A_{\mu\nu}
    \label{eq:barGamma}
\end{align}

%Of course, under a rotation, we also find:
%\begin{align}
%\bm{P}_N\rightarrow \bm{R}\bm{P}_N \equiv \bar{\bm{P}}_N
%\end{align}
%where $\bar{\bm{P}}_N$ is the momentum after rotation.

Therefore, in the end, we can prove Eq. \ref{eq:final2} using Eqs. \ref{eq:barP}, \ref{eq:barGamma} and \ref{eq:Drot}. Indeed, the rescaling direction for momentum will be the same (relative to the molecular frame) for any molecular orientation.
Note that, according to Eqs. \ref{eq:etf},\ref{eq:barp}, and \ref{eq:barGamma},
\begin{align}
\boxed{\bm{\Gamma}^A_{\bar{\mu}\bar{\nu}} \left(\bm{RX}\right)=\bm{R}\bm{\Gamma}^A_{\mu\nu}\left(\bm{X}\right)}.
\end{align}

\subsection{Equivalence of a Lagrangian approach and the approach based on a rotating basis}\label{appendix:lagrangian}
Here we, will show that the results above in Eqs. \ref{eq:center_munu}-\ref{eq:gamma_v1_final} (which were found by calculating the derivative coupling in a rotating basis) can also be  achieved by minimizing a constrained Lagrangian whereby we seek the smallest ERFs that satisfy Eqs. \ref{eq:gamma_cond1} and \ref{eq:gamma_cond3}.  The relevant Lagrangian is of the form:
\begin{align}
    &\mathcal{L} = \sum_{A \mu \nu} \frac{1}{\zeta^{A}_{\mu\nu}}\bm{\Gamma}''^{A \top}_{\mu \nu}\bm{\Gamma}''^{A}_{\mu \nu} - \sum_{\mu \nu}\bm{\lambda}_{1\mu\nu}^{\top}\left(\sum_A\bm{\Gamma}''^{A}_{\mu \nu}\right) -\sum_{\mu \nu}\bm{\lambda}_{2\mu\nu}^{\top} \left(\sum_{A} \bm{X}_A \times \bm{\Gamma}''^{A}_{\mu \nu}-\bm{J}_{\mu \nu}\right)\label{eq:s_L}
\end{align}

We will now show that the solution to this constrained problem is:
\begin{align}
    \bm{X}_{\mu\nu}^0 &= \sum_A \zeta_{\mu\nu}^A\bm{X}_A /\sum_A \zeta_{\mu\nu}^A\label{eq:s_R0}\\
    \bm{K}_{\mu\nu}&=-\sum_{A}\zeta_{\mu\nu}^A\left(\bm{X}_A-\bm{X}_{\mu\nu}^0\right)^\top\left(\bm{X}_A-\bm{X}_{\mu\nu}^0\right)\mathcal{I} + \sum_A\zeta_{\mu\nu}^A\left(\bm{X}_A-\bm{X}_{\mu\nu}^0\right)\left(\bm{X}_A-\bm{X}_{\mu\nu}^0\right)^\top\label{eq:s_K}\\
    \bm{\Gamma}''^{A}_{\mu\nu} &= \zeta_{\mu\nu}^A\left(\bm{X}_A-\bm{X}_{\mu\nu}^0\right) \times \left(\bm{K}_{\mu\nu}^{-1}\bm{J}_{\mu\nu}\right)\label{eq:s_gamma}
\end{align}

To being our derivation, note that the gradient of $\mathcal{L}$ in Eq. \ref{eq:s_L} w.r.t. $\bm{\Gamma}''^A_{\mu\nu}$ is zero, which reads
\begin{align}
    \frac{2}{\zeta_{\mu\nu}^A}\bm{\Gamma}''^A_{\mu\nu} - \bm{\lambda}_{1\mu\nu} - \bm{\lambda}_{2\mu\nu}\times \bm{X}_A = \bm{0}
\end{align}
For the simplicity of the notation, we may absorb the factor 2 into the Lagrangian multipliers and neglect the $\mu,\nu$ indices:
\begin{align}
    \bm{\Gamma}''^A = \zeta^A\bm{\lambda}_{1} + \zeta^A\bm{\lambda}_{2}\times \bm{X}_A
\end{align}
From the first constraint
\begin{align}
    \sum_A \bm{\Gamma}''^A = \bm{0}
\end{align}
we have
\begin{align}
    \bm{\lambda}_1 = -\bm{\lambda}_2\times\frac{\sum_A\zeta^A\bm{X}_A}{\sum_A\zeta^A} = -\bm{\lambda}_2\times\bm{X}^0
\end{align}
where $\bm{X}^0$ is defined in Eq. \ref{eq:s_R0}. Then
\begin{align}
    \bm{\Gamma}''^A = \zeta^A\bm{\lambda}_{2}\times \left(\bm{X}_A-\bm{X}^0\right)
\end{align}
Substituting the above equation into the second constraint,
\begin{align}
    \sum_{A} \bm{X}_A \times \bm{\Gamma}''^{A}=\bm{J}
\end{align}
we have
\begin{align}
    \sum_A \bm{X}_A\times\left(\zeta^A\bm{\lambda}_2\times\left(\bm{X}_A-\bm{X}^0\right)\right) = \bm{J}
\end{align}
the double cross product is
\begin{align}
    &\sum_A \bm{X}_A\times\left(\zeta^A\bm{\lambda}_2\times\left(\bm{X}_A-\bm{X}^0\right)\right)\nonumber\\
    =&\sum_A\zeta^A\bm{X}_A^\top\left(\bm{X}_A-\bm{X}^0\right)\bm{\lambda}_2 - \sum_A\zeta^A\left(\bm{X}_A-\bm{X}^0\right)\bm{X}_A^\top\bm{\lambda}_2\\
    =&\left(\left(\sum_A\zeta^A\bm{X}_A^\top\left(\bm{X}_A-\bm{X}^0\right)\right)\mathcal{I}-\sum_A\zeta^A\left(\bm{X}_A-\bm{X}^0\right)\bm{X}_A^\top\right)\bm{\lambda}_2
\end{align}
Let
\begin{align}
    \bm{K} = -\left(\sum_A\zeta^A\bm{X}_A^\top\left(\bm{X}_A-\bm{X}^0\right)\right)\mathcal{I}+\sum_A\zeta^A\left(\bm{X}_A-\bm{X}^0\right)\bm{X}_A^\top\label{eq:s_K_lag}.
\end{align}
We compute
\begin{align}
    \lambda_2 = -\bm{K}^{-1}\bm{J}
\end{align}
and therefore
\begin{align}
    \bm{\Gamma}''^A &=\zeta^A(-\bm{K}^{-1}\bm{J})\times \left(\bm{X}_A-\bm{X}^0\right)\\
    &=\zeta^A\left(\bm{X}_A-\bm{X}^0\right)\times(\bm{K}^{-1}\bm{J})
\end{align}
When recovering the $\mu,\nu$ indices, we have found:
\begin{align}
    \bm{\Gamma}''^A_{\mu\nu} = \zeta^A_{\mu\nu}\left(\bm{X}_A-\bm{X}^0_{\mu\nu}\right)\times(\bm{K}_{\mu\nu}^{-1}\bm{J}_{\mu\nu})
\end{align}
The only thing left is to show that $\bm{K}$ defined in Eq. \ref{eq:s_K_lag} is equivalent to Eq. \ref{eq:s_K}. Note that since
\begin{align}
    \sum_A \zeta^A\left(\bm{X}_A - \bm{X}^0\right) = \bm{0},
\end{align}
we can add $\bm{X}^{0\top}\left(\sum_A \zeta^A\left(\bm{X}_A - \bm{X}^0\right)\right)$ and $-\left(\sum_A \zeta^A\left(\bm{X}_A - \bm{X}^0\right)\right)\bm{X}^{0\top}$ to the first and second term of Eq. \ref{eq:s_K_lag}, respectively, which yields
\begin{align}
    \bm{K} = -\left(\sum_A\zeta^A\left(\bm{X}_A-\bm{X}^0\right)^\top\left(\bm{X}_A-\bm{X}^0\right)\right)\mathcal{I}+\sum_A\zeta^A\left(\bm{X}_A-\bm{X}^0\right)\left(\bm{X}_A-\bm{X}^0\right)^\top
\end{align}
which is exactly Eq. \ref{eq:s_K}. As a result, the solution to minimizing the Lagrangian in Eq. \ref{eq:s_L} is equivalent to Eqs. \ref{eq:s_R0}--\ref{eq:s_gamma}.
\subsection{Geometries for [5]Helicene and Methanol}\label{appendix:geometry}
\begin{table}[H]
\caption{Cartesian coordinates of [5]helicene in unit of Angstrom.}
\begin{tabular}{c c c c} 
\hline
C   &  1.169858 &  1.521282 & -0.904004 \\
C   &  2.022419 &  2.582079 & -1.139007 \\
C   &  3.350247 &  2.551820 & -0.667900 \\
C   &  3.812404 &  1.422047 & -0.027389 \\
C   &  2.968889 &  0.303892 &  0.186737 \\
C   &  1.588966 &  0.370167 & -0.186999 \\
C   &  0.723707 & -0.781432 &  0.054863 \\
C   &  1.367883 & -2.020903 &  0.328544 \\
C   &  0.650443 & -3.244646 &  0.207739 \\
C   & -0.649875 & -3.244715 & -0.207773 \\
C   & -1.367520 & -2.021066 & -0.328413 \\
C   & -0.723558 & -0.781522 & -0.054646 \\
C   & -1.588946 &  0.369996 &  0.187169 \\
C   & -2.968796 &  0.303559 & -0.187008 \\
C   & -3.812463 &  1.421610 &  0.026694 \\
C   & -3.350702 &  2.551392 &  0.667510 \\
C   & -2.023164 &  2.581731 &  1.139337 \\
C   & -1.170460 &  1.520877 &  0.904963 \\
H   & -0.162572 &  1.562077 &  1.297426 \\
H   & -1.666948 &  3.440348 &  1.702121 \\
H   & -4.012124 &  3.396249 &  0.838907 \\
H   & -4.849376 &  1.355036 & -0.294565 \\
C   & -3.509600 & -0.921295 & -0.687867 \\
C   & -2.752239 & -2.052082 & -0.689718 \\
H   & -3.185559 & -3.009065 & -0.970220 \\
H   & -4.554636 & -0.948962 & -0.986370 \\
H   & -1.176808 & -4.177601 & -0.391862 \\
H   &  1.177529 & -4.177477 &  0.391646 \\
C   &  2.752651 & -2.051731 &  0.689672 \\
C   &  3.509962 & -0.920863 &  0.687562 \\
H   &  4.554976 & -0.948444 &  0.986120 \\
H   &  3.186120 & -3.008619 &  0.970309 \\
H   &  4.849493 &  1.355555 &  0.293303 \\
H   &  4.011573 &  3.396656 & -0.839828 \\
H   &  0.161740 &  1.562284 & -1.295880 \\
H   &  1.665954 &  3.440783 & -1.701508 \\
\hline
\end{tabular} 
\end{table}
\begin{table}[H]
\caption{Cartesian coordinates of methanol in unit of Angstrom.}
\begin{tabular}{c c c c} 
\hline
C     & -0.652998   &  0.022929   & -0.000032 \\
O     &  0.736526   & -0.133385   &  0.000018 \\
H     & -0.980477   &  0.560041   & -0.916513 \\
H     & -0.980097   &  0.564936   &  0.913717 \\
H     & -1.127527   & -0.979878   &  0.002897 \\
H     &  1.113883   &  0.784406   & -0.000055 \\
\hline
\end{tabular} 
\end{table}

\bibliography{cite.bib}

\end{document}